\documentclass[a4paper,12pt]{article}
\usepackage{amsfonts,graphicx}
\newcommand{\rnc}{\renewcommand}
\rnc{\thesection}{\arabic{section}\setcounter{equation}{0}}
\rnc{\theequation}{\arabic{section}.\arabic{equation}}
\rnc{\thefootnote}{\alph{footnote}}
\newtheorem{conjecture}{Conjecture}[section]
\begin{document}
\title{Thermodynamic Bethe ansatz equation from 
fusion hierarchy of $osp(1|2)$ 
integrable spin chain}
\author{Kazumitsu Sakai\footnote{
e-mail address: sakai@printfix.physik.uni-dortmund.de; 
present address: Universit\"at Dortmund, 
Fachbereich Physik, Otto-Hahn-Str. 4 D-44221 Dortmund, Germany}  
 and Zengo Tsuboi\footnote{
JSPS Research Fellow; 
e-mail address: tsuboi@gokutan.c.u-tokyo.ac.jp} \\
\it c/o A. Kuniba, Institute of Physics,                                   
 University of Tokyo, \\
\it Komaba   3-8-1, Meguro-ku, Tokyo 153-8902, Japan}
\date{}
\maketitle
\begin{abstract}
The thermodynamic Bethe ansatz (TBA) 
and the excited state TBA equations for an 
integrable spin chain related to the Lie
superalgebra $osp(1|2)$ are proposed by the
quantum transfer matrix (QTM) method.
We introduce the fusion hierarchy of the QTM and 
derive the functional relations among them
($T$-system) and their certain combinations 
($Y$-system).
Their analytical property leads to the
non-linear integral equations which describe
the free energy and the correlation length at
any finite temperatures.
With regard to  the free energy, they coincide with the
TBA equation based on the string hypothesis.
\footnotetext{$^{*  \dagger}$alphabetical order}
\end{abstract}
Int. J. Mod. Phys. A, Vol. 15, p. 2329 - 2346 (2000)
\section{Introduction}
Solvable lattice models related to the Lie superalgebras \cite{Ka} 
have  attracted a great deal of attention 
\cite{KulSk,Kul,BS,KuR,DFI,Sa,ZBG,MR94}.  
For example, the supersymmetric $t-J$ model in  strongly 
correlated electron system has received much attentions 
in relation with the high $T_{c}$ superconductivity. 
 These models have both 
 fermionic and bosonic degree of freedom,
  and are given as solutions of the 
 graded Yang-Baxter equations \cite{KulSk}. 
To solve such models, the Bethe ansatz is widely used 
in many literatures 
(see, for example, \cite{Kul,Sch,EK,EKS,FK,EKtba,Ma,Mar95-1,
Mar95-2,RM,PF,MR} and references therein).  
However, many of them deal only with models related to 
 simple representations like fundamental ones; 
there was few {\em systematic} study by the Bethe ansatz 
on more complicated models such as 
fusion models \cite{KRS}. 

In view of such situations, we have 
recently carried out \cite{T1,T2,T3,T4,T5} 
{\em systematically} an 
analytic Bethe ansatz \cite{R1,R2,BR,S2,KS1,KS2,KOS} 
related to the Lie superalgebras 
$sl(r+1|s+1),B(r|s),C(s)$ and $D(r|s)$.  
Namely, we have proposed a class of 
 dressed vacuum forms (DVFs) labeled by Young (super) diagrams and 
 a set of fusion relations ($T$-system) among them. 

Besides the eigenvalue formulae of the transfer matrices, 
the thermodynamics have also been discussed by several people.  
 Particularly, the thermodynamic Bethe ansatz 
 (TBA) \cite{YY}
 equation was proposed for the supersymmetric $t-J$ model 
 \cite{Sch}, which is related to $sl(1|2)$, 
 and the supersymmetric extended 
 Hubbard model \cite{EKtba}, 
 which is related to $sl(2|2)$.  
 Moreover, there is a paper  \cite{JKSfusion} on  
 the excited state TBA equation  
 for these two important models of 
 1D highly correlated electron systems 
 from the point of view of the quantum transfer matrix 
 (QTM) method.  
 In addition, the TBA equation for $sl(r|s)$ model
  was presented \cite{Sa99} in 
 relation with the continuum limit of the integrable super spin chains.  
 
 However, the thermodynamics of the quantum spin model related to 
 the orthosymplectic Lie superalgebra $osp(r|2s)$ 
 is not so understood as $sl(r|s)$ case. 
 In particular, as far as we know, 
 there have been no literatures on the  
  TBA equation even for the simplest orthosymplectic 
 $osp(1|2)$ integrable spin chain 
 \cite{Kul,BS,KuR,Sa,ZBG,MR94,Mar95-1,Mar95-2}. 
 This is regrettable because this model may 
 be related to interesting topics such as 
 $N=1$ superconformal-symmetry in field theory, and 
 the loop model, which will describe 
 statistical properties of polymers 
 in condensed matter physics \cite{MNR}. 
 In view of these situation, we have recently 
 proposed the TBA equation for the $osp(1|2)$ integrable spin chain
 \cite{ST} 
 by using the string hypothesis \cite{G,T,TS}. 
 
Though we expect that the resultant TBA equation 
describes the free energy correctly, 
there exists the deviation from
the string hypothesis \cite{AlczMart,EKSstring,JuDo}. 
In addition, it is difficult 
to evaluate other physical quantities 
such as correlation length. 

As an alternative method which overcomes such
difficulties, the QTM method has been proposed
\cite{MSuzPB,InSuz,InSuz2,Koma,SAW,SNW}.  
Now we shall briefly sketch the QTM method. 
Utilizing the general equivalence theorem 
\cite{MSuzPr,MSuzPB}, one 
transforms the 1D quantum system into 
 the 2D classical counterpart and defines 
the QTM on such a fictitious system of size $N$ 
(referred to as the Trotter number, which should be 
taken $N\to\infty$). 
Since the QTM has a finite gap,
the original problem for the calculation 
of the partition function reduces 
to finding the single
largest eigenvalue of the QTM.
To evaluate it actually, we utilize the 
underlying integrable structure, which 
 admits introduction
of the ``commuting"  QTM with a complex 
parameter $v$ 
\cite{Klu,KZeit,JK,JKStJ,FK99,SSSU}. 
Furthermore, we introduce some auxiliary 
functions including the QTM itself, 
which satisfy functional relations. 

We select these auxiliary functions such that
they are
{\it A}nalytic, {\it N}on{\it Z}ero 
and have {\it C}onstant asymptotics 
in appropriate strips on the complex $v$-plane (we call
this property ANZC).
Thus we can transform the functional
relations into the non-linear integral equations (NLIE)
which describe the free energy.
In these NLIE, we can take the
Trotter limit $N\to\infty$. 

Adopting a subset of fusion hierarchy as auxiliary functions, 
we  find that these NLIE are 
 equivalent to the TBA equation based on 
the string hypothesis
\cite{JKSfusion,Klu,KSS,JSuz1,Sakai}.
 In general, a set of fusion hierarchy 
 satisfies 
the functional relations called $T$-system, 
and this $T$-system is transformed to 
 the $Y$-system \footnote{For simple Lie algebras, 
general relations 
between $T$ and $Y$-system 
are given in \cite{KNS1} and \cite{KNS2} 
(see also \cite{KP2}).}.
By selecting them so that they have 
ANZC property in appropriate strips,
one derives the NLIE which will be
identical to the TBA equation.
Furthermore, considering the sub-leading
eigenvalues of the QTM, we can derive systematically
the ``excited state" TBA equations which
 provide the correlation length 
at any finite temperatures.

The purpose of this paper is to apply our recent results \cite{T5} 
 to the $osp(1|2)$ integrable spin chain, and to 
  construct the TBA equation and it's excited state version 
 from the point of view of the above-mentioned 
 QTM method. 
  We have also confirmed the fact that our TBA equation 
coincides with the one \cite{ST} 
from the string hypothesis.  
We believe that this paper yields a  basis of future studies 
of the thermodynamics  by the QTM method 
for more general models such as  
the $osp(r|2s)$ model. 

The layout of this paper is as follows.
In section 2 we formulate the $osp(1|2)$ integrable
spin chain at finite temperatures in terms of the
commuting QTM.
In section 3, to evaluate its eigenvalue,
we utilize the fusion hierarchy of the QTM.
It is given as a set of the dressed vacuum forms 
(DVFs) which is  a summation over  
tableaux labeled by a Young (super) diagram with one column. 
These DVFs and their certain combinations ($Y$-functions)
satisfy the $T$-system and the $Y$-system,
respectively.
The formulation of the DVFs and  
the functional relations 
are essentially independent of the vacuum part. 
Thus we can utilize  the results in Ref. \cite{T5} 
only by replacing the vacuum parts of the DVFs defined
in Ref. \cite{T5} with those of the QTM. 
Based on the analytical property of the DVFs and the
$Y$-functions, in section 4 we derive the non-linear 
integral equations 
for the free energy, and in section 5 for the correlation 
length.
Section 6 is devoted to the summary and the discussion. 
\section{Quantum transfer matrix}
In this section, we introduce the quantum spin chain
related to the Lie superalgebra $osp(1|2)$ and
formulate the commuting QTM for this model.

The classical counterpart of this model is
given by the $\check{R}$-matrix (see for example, Ref. \cite{MR94})
\begin{eqnarray}
\check{R}(v)=I+vP^{g}-\frac{v}{v-\frac{3}{2}}E, \label{rmat}
\end{eqnarray}
which is a rational solution of the graded Yang-Baxter
equation \cite{KulSk} associated with the 
three dimensional representation of the Lie superalgebra 
$osp(1|2)$ (to be precise, ``super Yangian" $Y(osp(1|2))$), 
whose basis is ${\mathbb Z}_{2}$-graded and
labeled by the parameters $p(1)=p(\overline{1})=1$, $p(0)=0$. 
Here $P^{g}$ denotes the graded permutation 
operator which has $9\times 9$ matrix elements 
$ (P^{g})_{ab}^{cd}=(-1)^{p(a)p(b)} \delta_{a,d}\delta_{b,c}$ 
and $E$ is a $9\times 9$ matrix whose matrix elements are 
$ E_{ab}^{cd}=\alpha_{ab} (\alpha^{-1})_{cd}$ where 
 $\alpha $ is a $3 \times 3$ matrix
\begin{eqnarray}
\alpha=
\left(
 \begin{array}{@{\,}ccc@{\,}}
  \alpha_{11}  & \alpha_{10} & \alpha_{1\overline{1}} \\ 
  \alpha_{01}  & \alpha_{00} & \alpha_{0\overline{1}} \\ 
  \alpha_{\overline{1}1}  & \alpha_{\overline{1}0} & 
  \alpha_{\overline{1}\overline{1}} 
 \end{array}
\right)
=
\left(
 \begin{array}{@{\,}ccc@{\,}}
  0  & 0 & 1 \\ 
  0  & 1 & 0 \\ 
  -1 & 0 & 0 
 \end{array}
\right), \label{jirou}
\end{eqnarray}
and $\{a,b,c,d\}\in\{1,0,\overline{1}\}$
with the total order $1\prec 0\prec \overline{1}$.
The row-to-row transfer matrix $T(v)$ 
is defined by 
\begin{equation}
T(v)=\mbox{Tr}_a [R_{aL}(v)\cdots R_{a 1}(v)], \label{rtransf}
\end{equation}
where $L$ is the number of lattice sites and
$R_{aj}(v)$ denotes 
\begin{equation}
R(v)=P \check{R}(v),
\label{rmat2}
\end{equation}
which acts  non-trivially on 
the auxiliary space $a$ and the $j$-th site of the quantum space.
Note that $P_{ab}^{cd}=\delta_{a,d}\delta_{b,c}$ is 
the (non-graded) permutation operator.  
The Hamiltonian of the corresponding quantum system is given by 
taking the logarithmic derivative of above transfer matrix 
(\ref{rtransf}) at $v=0$,
\begin{eqnarray}
H=J\frac{d}{dv}\ln T(v) \biggl|_{v=0}
 =J\sum_{j=1}^{L}
 \left(
   P^{g}_{j, j+1}+\frac{2}{3}E_{j, j+1}
 \right) \label{hami}, 
\end{eqnarray}
where we assume a periodic boundary condition.
 Here $J$ is a real coupling constant 
which determines the phase of this model;
the ferromagnetic and antiferromagnetic regimes correspond 
$J>0$ and $J<0$, respectively (see for example, \cite{Mar95-2,ST}). 

To consider the finite temperature property 
of the model (\ref{hami}),
we shall introduce another transfer matrix 
$\widetilde{T}(v)$
constructed by  the $R$-matrix $\widetilde{R}(v)$, which is  
defined by $90^{\circ}$ rotation of $R(v)$, i.e.,
$\widetilde{R}_{jk}(v)={}^{t_k}\!R_{kj}(v)$ ($t_k$ means
the transposition for the $R$-matrix in the $k$-th space): 
\begin{equation}
\widetilde{T}(v)=
\mbox{Tr}_a [\widetilde{R}_{aL}(v)\cdots 
\widetilde{R}_{a 1}(v)]. \label{ltransf}
\end{equation}
We can see that the logarithmic derivative
of  above transfer matrix (\ref{ltransf})
also represents the Hamiltonian (\ref{hami}).
Thus the expansion of the transfer matrices (\ref{rtransf})
and (\ref{ltransf}) are expressed as 
\begin{eqnarray}
T(v)&=&T(0)\left\{1+\frac{H}{J}v+{\mathcal O}(v^2) \right\}, \nonumber \\
\widetilde{T}(v)&=&\widetilde{T}(0)
  \left\{1+\frac{H}{J}v+{\mathcal O}(v^2) \right\}.
\end{eqnarray}
Combining above relations and using the fact 
$T(0)\widetilde{T}(0)=1$ 
(note that $T(0)$ and $\widetilde{T}(0)$ denote
the right and left sift operators, respectively),
we have
\begin{equation}
T(v)\widetilde{T}(v)=1+\frac{2H}{J}v+{\mathcal O}(v^2).
\end{equation}
Consequently, the partition function $Z$ of the model (\ref{hami})
can be expressed as
\begin{equation}
Z=\mbox{Tr}e^{-\beta H}=\lim_{N\to \infty}\mbox{Tr}
  \left(T(u_N)\widetilde{T}(u_N)\right)^{N/2},\,\,u_N=-\frac{J\beta}{N},
\label{part}
\end{equation}
where $N$ is an even number:
it is called the Trotter number, 
which represents the number of fictitious sites in  the 
Trotter direction; $\beta=1/({k_B T})$ ($T$ is the
temperature).
Thus the free energy per site $f$ is given by
\begin{equation}
f=-\lim_{L\to\infty}\lim_{N\to\infty}\frac{1}{L\beta}\ln
   \mbox{Tr}\left(T(u_N)\widetilde{T}(u_N) \right)^{\frac{N}{2}}.
\label{free1}
\end{equation}
However, in the antiferromagnetic case $J<0$,
the eigenvalues of $T(u_N)\widetilde{T}(u_N)$
will be infinitely degenerate in the limit $N\to \infty$.
Therefore taking the trace in  above expression (\ref{free1}) 
is a serious problem.
To avoid this difficulty, we transform the term $T(u_N)\widetilde{T}(u_N)$
as follows: 
\begin{eqnarray}
\mbox{Tr}\left(T(u_N)\widetilde{T}(u_N) \right)^{\frac{N}{2}}&=&
\mbox{Tr}\prod_{k=1}^{N/2}\mbox{Tr}_{a_{2k},a_{2k-1}}\Bigl(
R_{a_{2k},L}(u_N)\cdots R_{a_{2k},1}(u_N) \nonumber \\
&& \times \widetilde{R}_{a_{2k-1},L}(u_N)\cdots  
\widetilde{R}_{a_{2k-1},1}(u_N)\Bigr) \nonumber \\
&=&\mbox{Tr}\prod_{j=1}^{L}\mbox{Tr}_j\prod_{k=1}^{N/2}
R_{a_{2k},j}(u_N)\widetilde{R}_{a_{2k-1},j}(u_N).
\end{eqnarray}
Now we introduce the QTM, which plays the fundamental
role to describe the thermal quantities 
\begin{equation}
T^{(1)}_1(u,v)=\mbox{Tr}_j \prod_{m=1}^{N/2}
R_{a_{2m},j}(u-iv)\widetilde{R}_{a_{2m-1},j}(u+iv).
\end{equation}
Due to the  Yang-Baxter equation,
we   see that the QTM is commutative:
\begin{equation}
[T^{(1)}_1(u,v),T^{(1)}_1(u,v^{\prime})]=0.
\end{equation}

Hereafter we write the $k$-th largest eigenvalue
of the QTM $T^{(1)}_1(u,0)$ as $T^{(1)}_{1,k}(u,0)$.
Since the two limits in (\ref{free1}) will be exchangeable
as proved in \cite{MSuzPB,InSuz}, we shall take
the limit $L\to\infty$ first.
Since there is a finite gap between 
$T^{(1)}_{1,1}(u_N,0)$
and $T^{(1)}_{1,2}(u_N,0)$ even in the
Trotter limit $N\to\infty$, we have
\begin{equation}
f=-\frac{1}{\beta}\lim_{N\to\infty}\ln T^{(1)}_{1,1}(u_N,0).
\label{free2}
\end{equation}
The thermodynamical completeness $-\lim_{\beta\to0}\beta f=
\ln 3$ follows from $T^{(1)}_{1,1}(0,0)=3$, which
is clear from $R_{12}(0)=P_{12}$.

Taking the ratio of the largest eigenvalue
and the second largest one, we can systematically 
calculate the correlation length $\xi$:
\begin{equation}
\frac{1}{\xi}=-\lim_{N\to\infty}\ln
\left|\frac{T^{(1)}_{1,2}(u_N,0)}{T^{(1)}_{1,1}(u_N,0)}\right|.
\label{corr}
\end{equation} 
%
\section{Fusion hierarchy and functional relations}
In this section, we present  the eigenvalue formulae of the 
fusion QTM $T_{m}^{(1)}(u,v)$ 
\footnote{We also use the expression $T_{m}^{(1)}(u,v)$
as the eigenvalues of the fusion QTM.} 
and the functional relations 
among them by using the results for 
the row-to-row transfer matrix \cite{T5}. 
Although the quantity we want to evaluate is only $T_{1}^{(1)}(u,0)$, 
 consideration on all $\{T_{m}^{(1)}(u,v)\}$ 
plays an essential role in our formulation.  
%
%
For $a \in \{1,0,\overline{1} \} $, we define
 the function $\framebox{$a$}_{v}$ as 
\begin{eqnarray} 
\hspace{-20pt} &&  
\framebox{$1$}_{v}=\psi_{1}(v)   
 \frac{Q(v-\frac{i}{2})}{Q( v+\frac{i}{2})}, \nonumber \\
\hspace{-20pt} && \framebox{$0$}_{v}=
   \psi_{0}(v)  
   \frac{Q(v)Q(v+\frac{3i}{2})}{Q(v+\frac{i}{2})Q(v+i)},
      \label{box}   \\
\hspace{-20pt} && 
\framebox{$\overline{1}$}_{v}=
 \psi_{\overline{1}}(v)
 \frac{Q(v+2i)}{Q(v+i)}, \nonumber 
\end{eqnarray}
where $Q(v)=\prod_{j=1}^{n}(v-v_{j})$; 
$n \in \{0,1,\dots, N \}$ 
 is a quantum number; $v_{j} \in {\mathbb C}$. 
 The vacuum parts of the functions 
$\framebox{$a$}_{v}$ (\ref{box}) are given as follows: 
\begin{eqnarray}
 \psi_{1}(v)&=&(-1)^{N-n}
     \frac{\phi_{+}(v) \phi_{-}(v+i) \phi_{+}(v-\frac{i}{2})}
     {\phi_{+}(v-\frac{3i}{2})}, \nonumber \\ 
 \psi_{0}(v)&=& \phi_{+}(v)\phi_{-}(v) 
   , \label{vacuum} \\
 \psi_{\overline{1}}(v)&=&(-1)^{N-n}
    \frac{\phi_{-}(v)\phi_{+}(v-i)\phi_{-}(v+\frac{i}{2})}
     {\phi_{-}(v+\frac{3i}{2})}, 
   \nonumber 
\end{eqnarray}
where 
\begin{eqnarray}
 \phi_{\pm}(v)=(v \pm i u)^{\frac{N}{2}}. \label{pm}
\end{eqnarray}  
The eigenvalue formula of the QTM is given 
as the summation over these boxes (\ref{box}): 
\begin{eqnarray}
 T_{1}^{(1)}(u,v)=\framebox{$1$}_{v}+
 \framebox{$0$}_{v}+\framebox{$\overline{1}$}_{v}. 
 \label{pikata}
\end{eqnarray}
 We have
 \footnote{
 Here $Res_{v=a}f(v)$ denotes the residue of a function $f(v)$ at $v=a $.
 } 
\begin{eqnarray}
\hspace{-45pt}&& Res_{v=-\frac{i}{2}+v_{k}}
 (\framebox{$1$}_{v}+\framebox{$0$}_{v})=0 ,
      \nonumber \\
\hspace{-45pt}&& Res_{v=-i+v_{k}}
(\framebox{$0$}_{v}+\framebox{$\overline{1}$}_{v})=0
      \label{res}, 
\end{eqnarray}
 under the following type of the Bethe ansatz equation (BAE): 
\begin{eqnarray}
&& \frac{\phi_{-}(v_{k}+\frac{i}{2})\phi_{+}(v_{k}-i)}
     {\phi_{-}(v_{k}-\frac{i}{2})\phi_{+}(v_{k}-2i)}
= 
 -(-1)^{N-n}
\frac{Q(v_{k}-\frac{i}{2})Q(v_{k}+i)}
     {Q(v_{k}+\frac{i}{2})Q(v_{k}-i)} \nonumber \\ 
&& \hspace{150pt} {\rm for} \quad k \in \{1,2,\dots,n\} 
       \label{BAE}.
\end{eqnarray}
Thus $T_{1}^{(1)}(u,v)$ (\ref{pikata}) is free
 of poles\footnote{Here singularities of the vacuum parts 
 of the DVFs, which can be removed by 
 multiplying overall vacuum functions
  are out of the question.}
 under this BAE (\ref{BAE}). 
 We observe, from numerical analysis, 
 that the largest and the second largest eigenvalues 
  lie in the sector $n=N$ and $n=N-1$, respectively. 

 We shall introduce a function 
 $T^{(1)}_{m}(u,v)$ 
with a spectral parameter $v\in {\mathbb C}$ and 
 a Young (super) diagram
\footnote{
Note that this diagram corresponds to the 
Kac-Dynkin label $2m$.}
 $(1^{m})$, 
 which is a candidate of the DVF for the fusion QTM. 
 From now on, we often abbreviate the 
 parameter $u$ to simplify the notation.
For $m \in {\mathbb Z}_{\ge 1}$, 
we define  $T^{(1)}_{m}(v)$ 
as a summation 
 over products of 
the boxes in (\ref{box}) : 
\begin{eqnarray}
T^{(1)}_{m}(v) &=&
 \frac{1}{{\mathcal N}_{m}(v)} 
 \sum_{\{i_{k}\}\in B(1^{m})}
  \begin{array}{|c|}\hline 
     	i_{1}  \\ \hline
        i_{2}  \\ \hline
     	\vdots \\ \hline 
     	i_{m}  \\ \hline
  \end{array},
  \label{DVF-tableau} 
\end{eqnarray}
where the spectral parameter $v$ is shifted as 
$v+\frac{m-1}{2i},v+\frac{m-3}{2i},\dots ,v+\frac{-m+1}{2i}$ 
from the top to the bottom. 
$B(1^{m})$ is a set of tableaux
 $\{i_{k}\}$ ($i_{k} \in \{1,0,\overline{1} \}$)
obeying the following rule 
(admissibility conditions) 
\begin{equation}
 i_{1} \preceq i_{2} 
   \preceq \cdots \preceq i_{m}.
\end{equation}
The normalization function ${\mathcal N}_{m}(v)$ is introduced 
so that the degree of $T^{(1)}_{m}(v)$ is $2N$ with respect to $v$: 
\begin{eqnarray}
{\mathcal N}_{m}(v)=\prod_{j=1}^{m-1} 
 \phi_{-}\left(v+\frac{m+1-2j}{2}i\right)
 \phi_{+}\left(v-\frac{m+1-2j}{2}i\right).
\end{eqnarray}
The DVF (\ref{DVF-tableau}) 
is 
free of poles under the BAE (\ref{BAE})
 for any $m \in{\mathbb Z}_{\ge 1}$.  
Note that $T^{(1)}_{m}(v)$ (\ref{DVF-tableau}) for $m=1$ 
reduces to (\ref{pikata}) since ${\mathcal N}_{1}(v)=1$. 
The BAE (\ref{BAE}) for the QTM has the same form as the one 
\cite{Kul,Mar95-1,Mar95-2,MR} 
for the row to row transfer matrix except the vacuum part.  
Thus the dress part of the function $T_{m}^{(1)}(v)$ 
(\ref{DVF-tableau}) 
 is same as that in \cite{T5}.  
Therefore the DVFs $T_{m}^{(1)}(v)$ ($m \in {\mathbb Z}_{\ge 1}$) 
satisfy the following functional relations
\footnote{
Of course, there is a large class of functional relations 
among DVFs labeled by any skew-Young diagrams. 
The functional relations, which we are now dealing with  
is nothing but a subset of it. 
However we know, from our 
experience, that the functional relations among DVFs labeled by 
{\it rectangular} Young diagram often have a good 
analyticity relevant for the TBA. 
The $T$-system (\ref{T-sys}) is one of them. 
}  
($osp(1|2)$ version of the fusion relations
 or the $T$-system
\footnote{In this paper, 
we use the variable $m$ in (\ref{T-sys}) slightly 
 different from that in \cite{T5} for practical reason: 
 the variable $m$ in (\ref{T-sys}) 
corresponds to $2m$ in \cite{T5}. 
}
), 
which has essentially the same form in \cite{T5}:
\begin{eqnarray} 
\hspace{-40pt}
&& T_{m}^{(1)}\left(v-\frac{i}{2}\right)
T_{m}^{(1)}\left(v+\frac{i}{2}\right)
=T_{m-1}^{(1)}(v)T_{m+1}^{(1)}(v)+
T_{m}^{(0)}(v)T_{m}^{(1)}(v), \label{T-sys}
\end{eqnarray}
where 
\begin{eqnarray} 
\hspace{-40pt}
&& T_{0}^{(1)}(v)=\phi_{-}\left(v+\frac{i}{2}\right)
\phi_{+}\left(v-\frac{i}{2}\right), 
  \nonumber \\ 
&& T_{m}^{(0)}(v)=
  \frac{\phi_{-}(v+\frac{m+2}{2}i)\phi_{+}(v-\frac{m+2}{2}i)
        \phi_{-}(v-\frac{mi}{2})\phi_{+}(v+\frac{mi}{2})}
       {\phi_{-}(v+\frac{m+1}{2}i)\phi_{+}(v-\frac{m+1}{2}i)}
       \nonumber \\ 
&& \hspace{170pt} {\rm for} \qquad m \in {\mathbb Z}_{\ge 1}.
     \label{T0} 
\end{eqnarray}
We can prove these functional relations by the 
Jacobi identity and a duality among the DVFs \cite{T5}. 

Using $T_{m}^{(1)}(v)$ (\ref{DVF-tableau}), 
we introduce the following functions ($Y$-functions):
\begin{eqnarray} 
Y_{m}(v):=\frac{T_{m-1}^{(1)}(v)T_{m+1}^{(1)}(v)}
               {T_{m}^{(0)}(v)T_{m}^{(1)}(v)} 
           \quad {\rm for} \quad m \in {\mathbb Z}_{\ge 1}. 
                \label{Y-fun}
\end{eqnarray}
These functions (\ref{Y-fun}) satisfy a so-called $Y$-system:
\begin{eqnarray} 
Y_{m}(v-\frac{i}{2})Y_{m}(v+\frac{i}{2})
 =\frac{(1+Y_{m-1}(v))(1+Y_{m+1}(v))}
       {1+\{Y_{m}(v)\}^{-1}} 
  \quad {\rm for} \quad m \in {\mathbb Z}_{\ge 1}, 
                \label{Y-sys}
\end{eqnarray}
where $T_{-1}^{(1)}(v)=Y_{0}(v)=0$.  
In the subsequent sections, these functional relations (\ref{Y-sys}) 
will be transformed into the TBA equations. 
\section{Thermodynamic Bethe ansatz equation}
%
Let $\{T^{(1)}_{m,k}(u,v)\}$ be the DVFs 
$\{T^{(1)}_{m}(u,v)\}$ constructed by the 
 roots of BAE (\ref{BAE}) which provide  the 
$k$-th largest eigenvalue of the QTM $T_{1}^{(1)}(u,v)$, 
and $\{Y_{m,k}(u,v)\}$ be the $Y$-functions constructed by 
$\{T_{m,k}^{(1)}(u,v)\}$ as in (\ref{Y-fun}).

\begin{figure}
\includegraphics[width=0.95\textwidth]{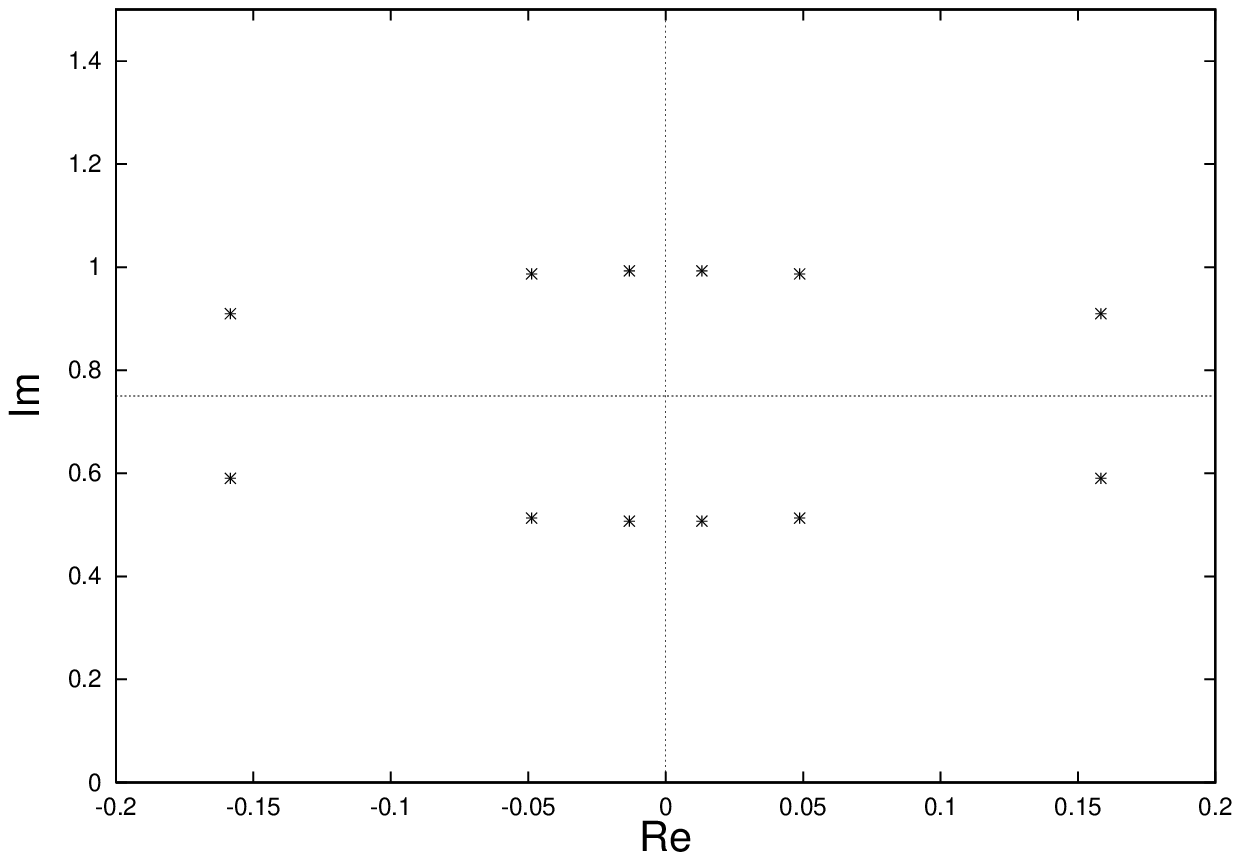}
\caption{The distribution  of the BAE roots
which  provides 
the largest eigenvalue for $N=12$, $u=0.05$.
There are six two-string solutions which are
symmetric with respect to the imaginary axis and the line
$\Im v=3/4$.}
\label{roots}
\end{figure}

In this section, we study the analyticity of the DVFs 
$\{T^{(1)}_{m,1}(u,v)\}$ 
and the $Y$-functions $\{Y_{m,1}(u,v)\}$ in the complex $v$-plane.
Then we derive the non-linear integral equations (NLIE) 
which provide the free energy.
For this purpose, by keeping the Trotter number $N$ finite,
we have performed numerical analyses 
with various values of $\beta$ and $N$ in investigating the 
location of zeros of $\{T^{(1)}_{m,1}(u,v)\}$.
As seen in Fig.\ref{roots},
there are $N/2$ two-string solutions of the BAE (\ref{BAE}),
which will provide the largest eigenvalue of the QTM.
These roots locate symmetrically with respect to the
imaginary axis and the line $\Im v=3/4$.

\begin{figure}
\includegraphics[width=0.95\textwidth]{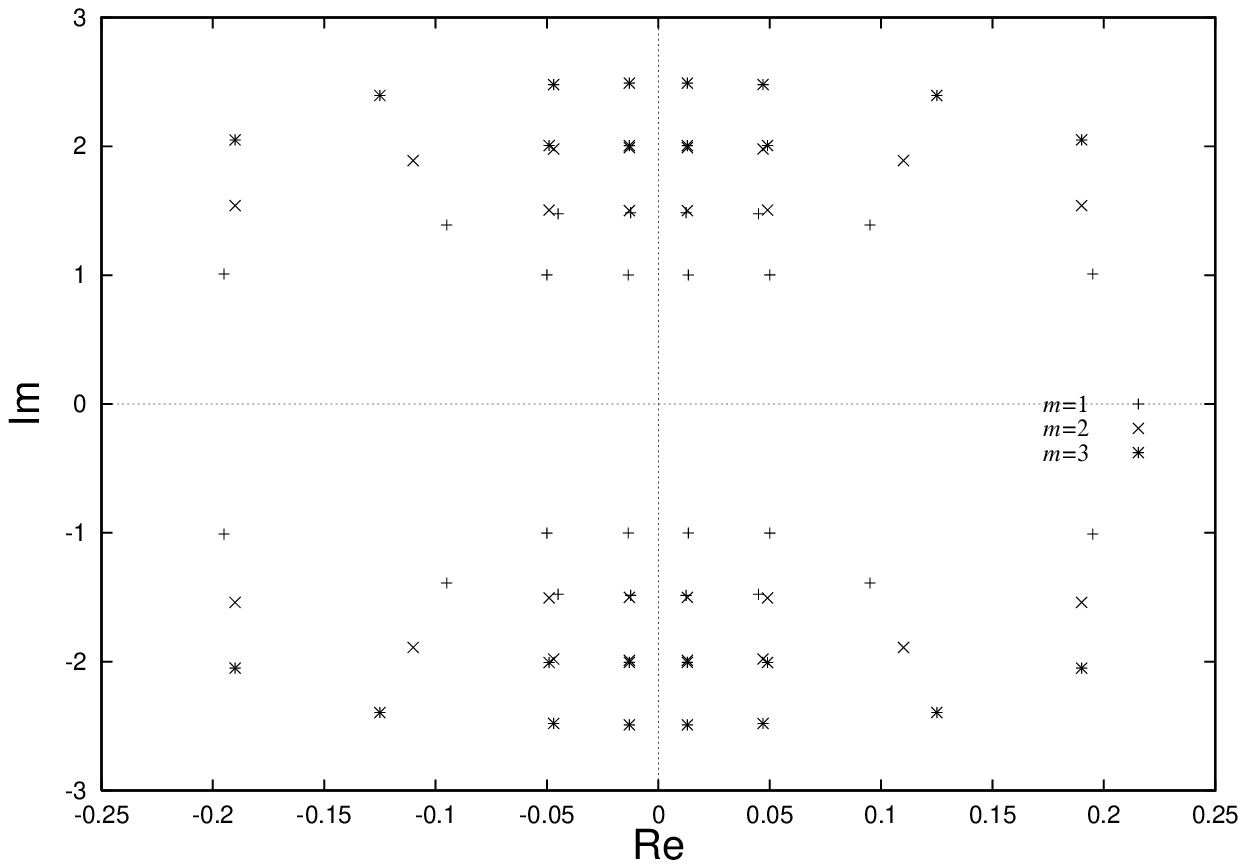}
\caption{Location of zeros for $T^{(1)}_{m,1}(u,v)$
for $m=1,2,3$, $u=0.05$ and $N=12$. 
Note that these zeros are distributed symmetrically with respect to 
the real and the imaginary axis.}
\label{zeroslgt12}
\end{figure}

{}From the definition of the DVFs
(\ref{DVF-tableau}), they have $2N$ zeros on the complex $v$-plane.
For example, the location of zeros of $T^{(1)}_{m,1}(u,v)$
for $u=0.05$, $N=12$ is plotted in Fig.\ref{zeroslgt12}.
According to them, we observe that
$T_{m,1}^{(1)}(u,v)$ have $N$ zeros on 
the smooth curve near the line
$\Im v = \pm (m+1)/2$ and
$N$ zeros  on the smooth curve near the line
$\Im v = \pm (m/2+1)$. 
Thus we expect that the following conjecture is valid
even in the limit $N\to \infty$.
\begin{conjecture} \label{conj-lar}
All the zeros of $T_{m,1}^{(1)}(u,v)$ 
for small $|u|\ll 1$ are located outside
of the strip $\Im v\in [-1/2,1/2]$. 
\end{conjecture}
{}From the definition (\ref{DVF-tableau}), one observes the fact
that all the zeros are symmetric 
with respect to both  real and imaginary axes.
As seen in the figure, the deviation from the
line is very small, which will be smaller as $u\to 0$.

Once this conjecture is assumed, we can identify the
strip where the $Y$-functions $\{Y_{m,1}(u,v)\}$
have the property of {\it A}nalytic {\it N}on{\it Z}ero 
and {\it C}onstant
asymptotics in the limit $v\to\infty$ (we call it
ANZC property).
Using the definition of the DVFs (\ref{DVF-tableau})
and the $Y$-functions (\ref{Y-fun}), we can obtain the asymptotic
value of the $Y$-functions in the limit $|v| \to\infty$:
\begin{equation}
\lim_{|v| \to \infty}Y_{m,1}(v)=\frac{m(m+3)}{2}.
\label{yasympt}
\end{equation}
{}From conjecture \ref{conj-lar} 
 and  above asymptotics (\ref{yasympt}),
the functions $1+Y_{m,1}(u,v)$ and $1+\{Y_{m,1}(u,v)\}^{-1}$
have ANZC property in the strip $\Im 
v\in[-\varepsilon,\varepsilon]$
($0< \varepsilon \ll 1$).  
On the other hand, the functions $Y_{m,1}(u,v)$ have the ANZC property
in the strip $\Im v\in[-1/2,1/2]$ (we call this strip
physical strip), except for $Y_{1,1}(u,v)$ possesses
poles (resp. zeros) of order $N/2$ at $\pm(1/2+u)i$
(resp. $\pm(1/2-u)i$) in the physical strip 
for $J>0$ (resp. $J<0$). Note that $u=u_N$ is
a small quantity given in (\ref{part}).
Using Cauchy's theorem, 
we can transform the $Y$-system
(\ref{Y-sys}) into the NLIE
in the following way.
Let us consider the $Y$-system (\ref{Y-sys}) for $m\ge 2$.
First, we take the logarithmic derivative and
perform the Fourier transformation on both side of
the Eq.(\ref{Y-sys}).
Second, by Cauchy's theorem, the Fourier integral for the logarithmic
derivative of $Y_{m,1}(v)$ is represented by that of
$1+Y_{m\pm 1,1}(v)$ and $1+\{Y_{m,1}(v)\}^{-1}$. 
Finally, performing the inverse Fourier transformation and integrating
over $v$, we obtain the desired NLIE. 
The integral constants are determined by the asymptotic value 
in (\ref{yasympt}). 
In the case $m=1$ in (\ref{Y-sys}), 
we have to modify $Y_{1,1}(v)$ 
to transform $Y$-system into the NLIE 
since $Y_{1,1}(v)$ have the poles (or zeros) in the physical strip. 
\setcounter{equation}{1}
\begin{equation}
\widetilde{Y}_{1,1}(v)=Y_{1,1}(v)\{
\tanh\frac{\pi}{2}(v+i(1/2\pm u))\tanh\frac{\pi}{2}
(v-i(1/2\pm u))\}^{\pm N/2},
\label{modY}
\end{equation}
where the $+$ and $-$ signs in front of $u$ and $N/2$ 
should be chosen according as $J>0$ and $J<0$,
respectively.
According to the identity $\tanh\frac{\pi}{4}(v+i)\tanh\frac{\pi}{4}
(v-i)=1$, $Y_{1,1}(v\pm i/2)$ in the lhs of the Eq.(\ref{Y-sys})
can be replaced by $\widetilde{Y}_{1,1}(v)$ in 
(\ref{modY}).   
Thus the $Y$-system (\ref{Y-sys}) can be transformed
to the NLIE in the similar way as mentioned above.
Consequently the resultant NLIE are represented as
\begin{eqnarray}
\ln Y_{1,1}(v)&=&
\mp \frac{N}{2}\ln\left\{\tanh\frac{\pi}{2}\left(
v+i(1/2\pm u)\right)\tanh\frac{\pi}{2}\left(
v-i(1/2 \pm u)\right)\right\} \nonumber \\
&&+K*\ln(1+Y_{2,1})(v)-
K*\ln(1+Y_{1,1}^{-1})(v), \nonumber \\
\ln Y_{m,1}(v)&=&K*\ln(1+Y_{m+1,1})(v)+K*\ln(1+Y_{m-1,1})(v) \nonumber \\
&&-K*\ln(1+Y_{m,1}^{-1})(v) \qquad \mbox{for } m\in{\mathbb Z}_{\ge 2},
 \label{goro} 
\end{eqnarray}
where $A*B(v)$ denotes the convolution 
\begin{equation}
A*B(v)=\int_{-\infty}^{\infty}A(v-v^{\prime})B(v^\prime)
dv^\prime,
\end{equation}
and 
\begin{equation}
K(v)=\frac{1}{2\cosh\pi v}.
\label{kernel}
\end{equation}
In  above NLIE (\ref{goro}), the Trotter limit ${N\to\infty}$ can
be calculated:
\begin{eqnarray}
&& \mp 
\lim_{N \to \infty}
\frac{N}{2}\ln\left\{\tanh\frac{\pi}{2}\left(
v+i(1/2 \pm u_N)\right)\tanh\frac{\pi}{2}\left(
v-i(1/2 \pm u_N)\right)\right\}
\nonumber \\ 
\hspace{30pt} &&=\frac{\pi \beta J}{\cosh\pi v}.
\end{eqnarray}
We thus arrive at the NLIE for $Y_{m,1}(v)$ which 
are independent of the Trotter number $N$.
\begin{eqnarray}
\ln Y_{1,1}(v)&=&\frac{\pi \beta J}{\cosh\pi v} 
+K*\ln(1+Y_{2,1})(v)-
K*\ln(1+Y_{1,1}^{-1})(v), \nonumber \\
\ln Y_{m,1}(v)&=&K*\ln(1+Y_{m+1,1})(v)+K*\ln(1+Y_{m-1,1})(v) \nonumber \\
&&-K*\ln(1+Y_{m,1}^{-1})(v) \qquad \mbox{for } m\in{\mathbb Z}_{\ge 2},
\label{tba}
\end{eqnarray}
Under the identification $Y_{m,1}(v)=\eta_m(v)$, above
equations (\ref{tba}) are identical to the 
TBA equation (30) in \cite{ST}, which was derived by the thermodynamic
Bethe ansatz based on the string hypothesis.
{}From above equations and the asymptotics (\ref{yasympt}), 
we can determine the $Y$-functions $Y_{m,1}(v)$ uniquely.

To obtain the free energy per site, we shall modify 
$T^{(1)}_{1,1}(v)$ as
\begin{equation}
\widetilde{T}_{1,1}^{(1)}(v)=\frac{T_{1,1}^{(1)}(v)}{\phi_{+}(v-i)
                         \phi_{-}(v+i)}.
\end{equation}  
{}From the definition of the $Y$-function (\ref{Y-fun}) 
 and the $T$-system  (\ref{T-sys})
$\widetilde{T}_{1,1}^{(1)}(v)$ satisfies the following
functional relation.
\begin{equation}
\frac{\widetilde{T}_{1,1}^{(1)}(v+\frac{i}{2})
               \widetilde{T}_{1,1}^{(1)}(v-\frac{i}{2})}
              {\widetilde{T}_{1,1}^{(1)}(v)}=F(v)(1+Y_{1,1}(v)), 
\end{equation}
where 
\begin{equation}
F(v)=\frac{\phi_{+}(v+\frac{i}{2})\phi_{-}(v-\frac{i}{2})}
              {\phi_{+}(v-\frac{i}{2})\phi_{-}(v+\frac{i}{2})}.
\end{equation}
ANZC property of the both sides leads to
\begin{eqnarray}
\ln T_{1,1}^{(1)}
(u_N,v)&=&G*\ln(1+Y_{1,1})(v)+\ln\phi_{+}(v-i)\phi_{-}(v+i) \nonumber \\
&& +N\int_{0}^{\infty}
     \frac{2e^{-\frac{k}{2}}\sinh ku_N \cos kv}
          {k(2\cosh\frac{k}{2}-1)}dk,
\end{eqnarray}
where $G(v)$ denotes the kernel,
\begin{equation}
G(v)=\frac{2}{\sqrt{3}}\frac{\sinh\frac{4}{3}\pi v}{\sinh 2\pi v}.
\end{equation}
Calculating the Trotter limit $N\to\infty$,
we obtain the free energy per site (\ref{free2})
\begin{equation}
f=J\left(\frac{4\pi}{3\sqrt{3}}-1\right)-k_BT\int_{-\infty}^
{\infty}G(v)\ln(1+Y_{1,1}(v))dv.
\label{largest}
\end{equation}
This representation coincides with Eq.(33) in \cite{ST}.
%
%
\section{Excited state TBA equation} 
\begin{figure}
\includegraphics[width=0.95\textwidth]{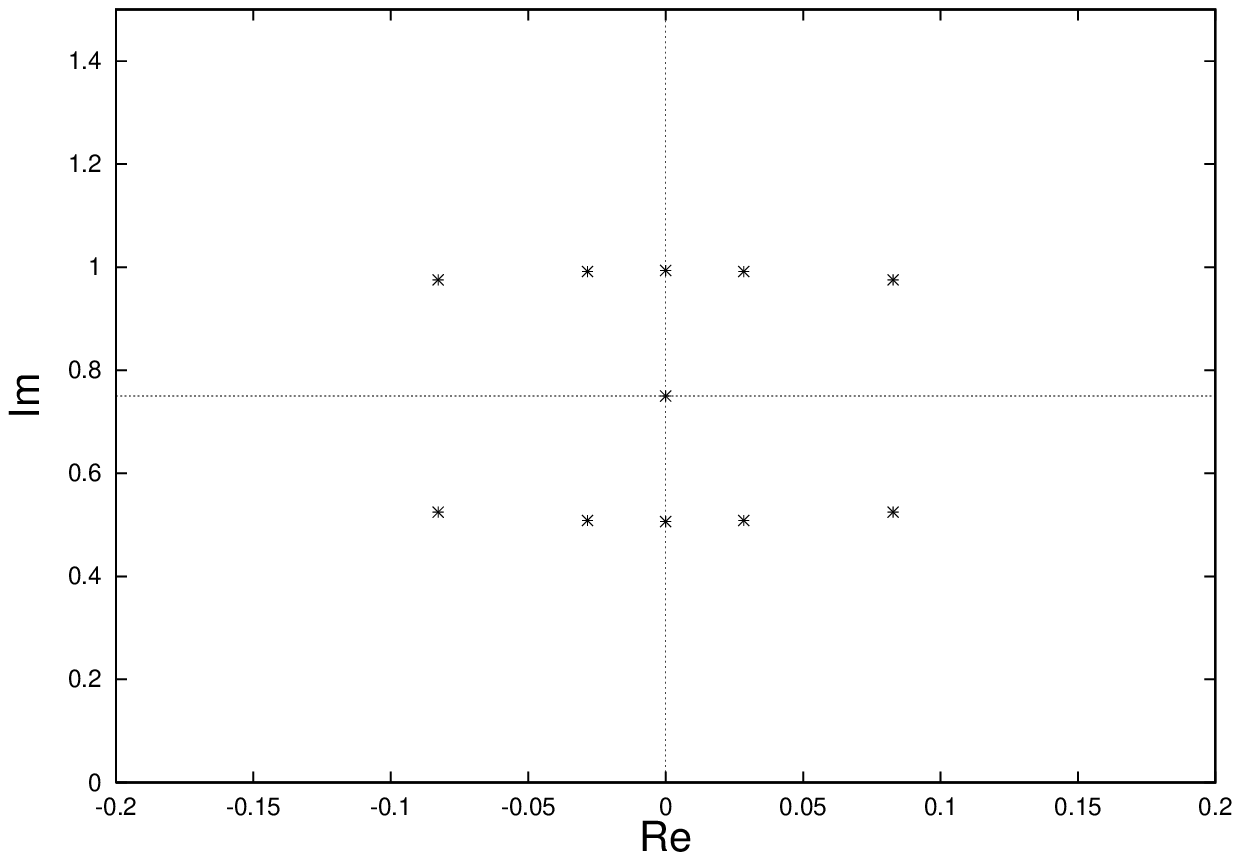}
\caption{Location of the roots of the BAE (\ref{BAE}) 
for  $u=0.05$, $N=12$ and $n=11$, which gives 
the second largest eigenvalue of the QTM. 
There are four two-strings and one three-string. 
They are symmetrically distributed with respect to 
the line $\Im v=3/4$ and the imaginary axis. }
\label{roots122nd}
\end{figure}

\begin{figure}
\includegraphics[width=0.95\textwidth]{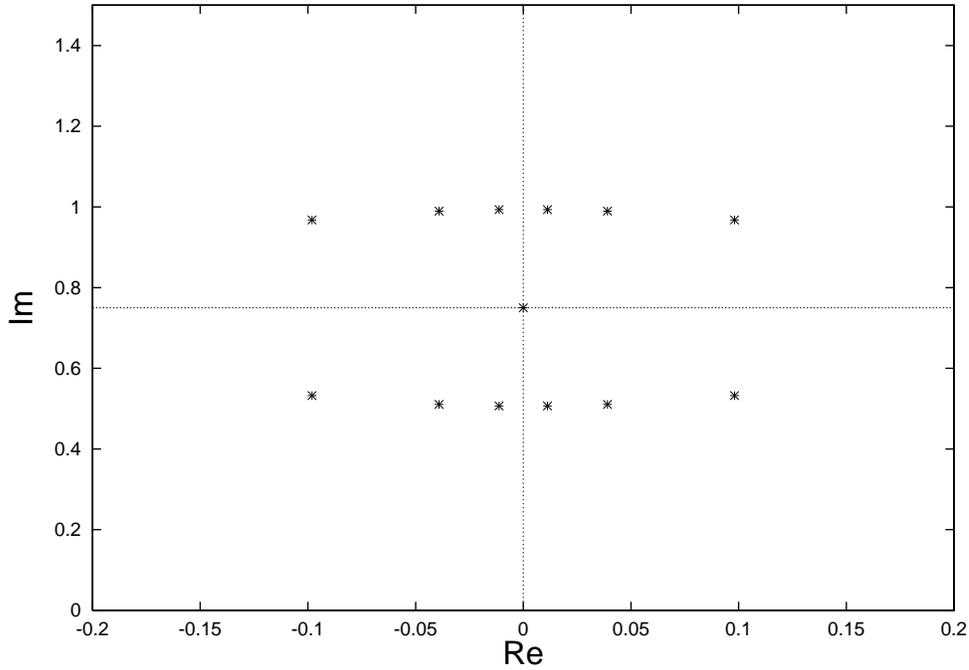}
\caption{Location of the roots of the BAE (\ref{BAE}) 
for  $u=0.05$, $N=14$ and $n=13$, which gives the  
second largest eigenvalue of the QTM. 
There are six two-strings and one one-string. 
They are symmetrically distributed with respect to 
the line $\Im v=3/4$ and the imaginary axis. }
\label{roots142nd}
\end{figure}

In this section, we develop the analysis 
for the largest eigenvalue of the QTM to 
the the analysis of the second largest one.  
Namely we derive the NLIE (excited state TBA equation) 
which provide the correlation length at finite temperature. 
We assume that the coupling constant  
$J$ is negative: we consider only the 
antiferromagnetic regime in this section. 
After some numerical analyses, 
we observe that $n=N-1$ roots $\{v_{k}\}$ 
of the BAE (\ref{BAE}), which will bring us the 
second largest eigenvalue of 
the QTM, form 
$\frac{N}{2}-2$ two-strings and one three-string if $N \in 4{\mathbb Z}$ 
(cf. Fig. \ref{roots122nd}); 
$\frac{N}{2}-1$ two-strings and one one-string if $N \in 4{\mathbb Z}+2$ 
(cf. Fig. \ref{roots142nd}). 
As in the largest eigenvalue case, 
the distribution of these roots are symmetric 
with respect to the imaginary axis and the line $\Im v=3/4$. 

\begin{figure}
\includegraphics[width=0.95\textwidth]{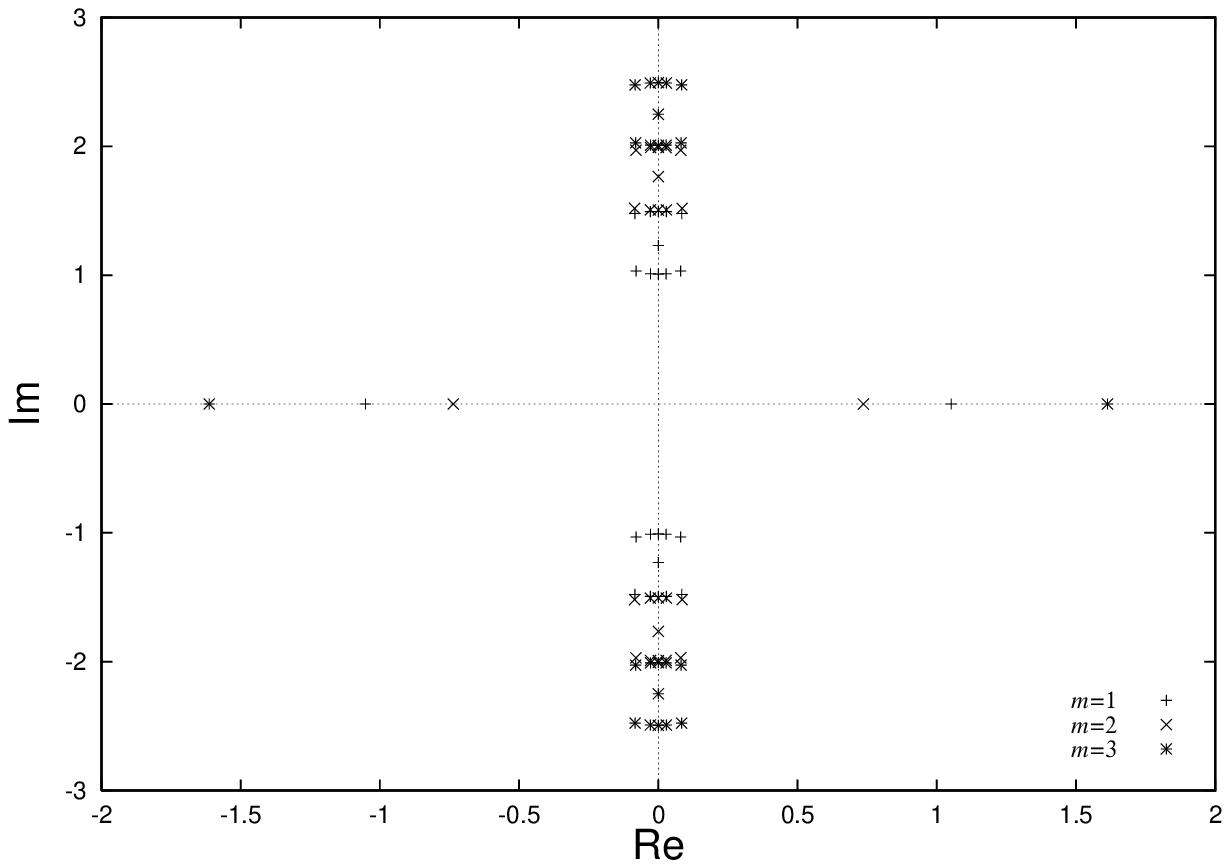}
\caption{Location of zeros of $T^{(1)}_{m,2}(u,v)$
for $m=1,2,3$, $u=0.05$ and $N=12$. 
Note that these zeros are distributed symmetrically with respect to 
the real and the imaginary axis. }
\label{zeroslgt122nd}
\end{figure}

\begin{figure}
\includegraphics[width=0.95\textwidth]{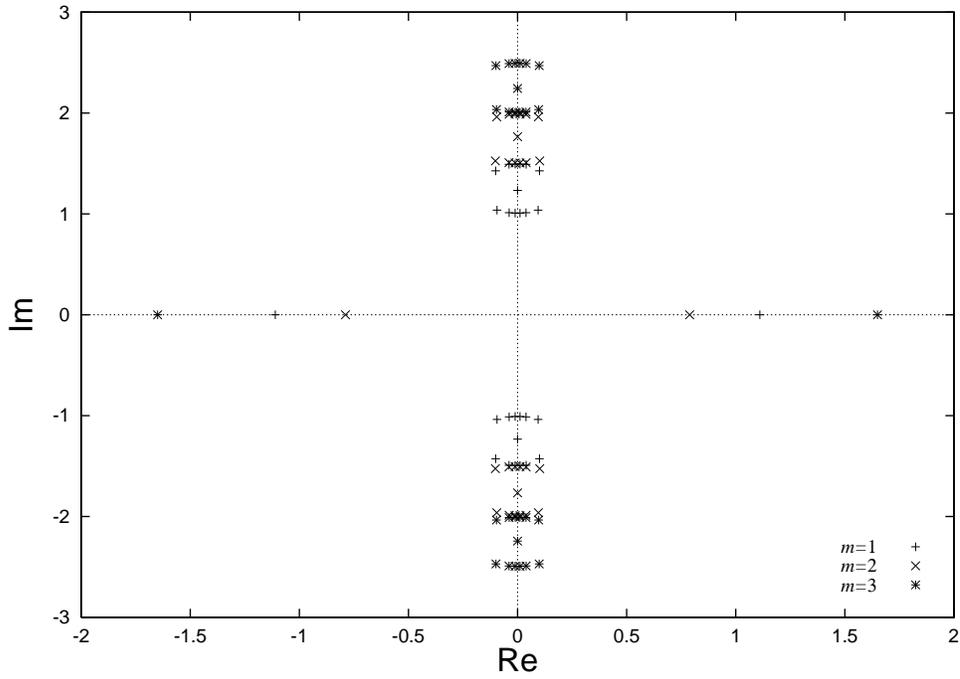}
\caption{Location of zeros of $T^{(1)}_{m,2}(u,v)$
for $m=1,2,3$, $u=0.05$ and $N=14$. 
Note that these zeros are distributed symmetrically with respect to 
the real and the imaginary axis.}
\label{zeroslgt142nd}
\end{figure}

From the definition of the DVF (\ref{DVF-tableau}), 
$T_{m,2}^{(1)}(u,v)$ has $2N$ zeros on the complex $v$-plain. 
From the numerical analyses, we observe that 
$T_{m,2}^{(1)}(u,v)$ has $N-2$ zeros on 
the smooth curve near the line $\Im v = \pm (m+1)/2$, 
$N-2$ zeros on the smooth curve near the line 
$\Im v = \pm (m/2+1)$, 
two zeros on the imaginary axis near the points  
$v = \pm (2m+3)i/4$ and 
two zeros $\pm x_{m}$ ($x_{m}>0$) on the real axis.  
For example, the location of zeros of $T^{(1)}_{m,2}(v)$ for 
$u=0.05$ and  $N=12,14$ are plotted in
 Figs. \ref{zeroslgt122nd},\ref{zeroslgt142nd}. 
We expect that the following conjecture is valid 
even in the limit $N \to \infty$. 
\begin{conjecture}\label{conj2}
$T_{m,2}^{(1)}(u,v)$ has two real zeros $\pm x_{m}$.  
Every other zero is located outside of the physical strip
$\Im v\in[-1/2,1/2]$.
\end{conjecture}

As in the previous section, we shall identify the strip 
where the $Y$-functions $ \{ Y_{m,2}(u,v) \} $ 
 have the ANZC property. 
From the definition of the DVFs (\ref{DVF-tableau}) and 
 the $Y$-functions (\ref{Y-fun}), 
we find the asymptotic value of $Y_{m,2}(v)$: 
\begin{equation}
\lim_{|v| \to \infty}Y_{m,2}(v)=\frac{(-1)^{m}(2m+3)-3}{4}.
\end{equation}
From the definition of (\ref{Y-fun}) and 
conjecture \ref{conj2}, 
we find $(1+Y_{m+1,2}(v))(1+Y_{m-1,2}(v))/(1+\{ Y_{m,2}(v) \}^{-1})$ has 
 the ANZC property in the strip 
 $ \Im v \in [-\varepsilon, \varepsilon]$ ($0<\varepsilon \ll 1$), 
 while $Y_{m,2}(v)$ has zeros of order 1 at $v=\pm x_{m+1},\pm x_{m-1}$ 
and poles of order 1 at $v=\pm x_{m}$ if $m \in {\mathbb Z}_{\ge 2}$; 
$Y_{1,2}(v)$ has 
zeros of order $N/2$ at $v=\pm (1/2-u)i$, 
zeros of order 1 at $v=\pm x_{2}$ 
and poles of order 1 at $v=\pm x_{1}$, 
and that every other singular point locates outside of the physical strip. 
Thus we have to modify  $Y_{m,2}(v)$  as follows: 
\setcounter{equation}{1}
\begin{eqnarray}
\widetilde{Y}_{m,2}(v)&=& Y_{m,2}(v)
 \left\{
\tanh\frac{\pi}{2}(v+i(\frac{1}{2}- u))\tanh\frac{\pi}{2}
(v-i(\frac{1}{2}- u))
\right\}^{-\frac{N\delta_{m,1}}{2}}
   \nonumber \\             
&&  \times
     \left\{\tanh\frac{\pi}{2}(v-x_{m})
            \tanh\frac{\pi}{2}(v+x_{m})\right\}
        \nonumber \\             
&&  \times  \left\{\tanh\frac{\pi}{2}(v-x_{m+1})
            \tanh\frac{\pi}{2}(v+x_{m+1})\right\}^{-1} 
        \nonumber \\             
&&  \times 
     \left\{\tanh\frac{\pi}{2}(v-x_{m-1})
            \tanh\frac{\pi}{2}(v+x_{m-1})\right\}^{-1+\delta_{m,1}} 
   \label{es-Y-fun}.    
\end{eqnarray}
These $\widetilde{Y}_{m,2}(v)$  (\ref{es-Y-fun}) 
have the ANZC property in the physical strip. 
After similar calculation in the previous section, 
we obtain the excited state TBA equation: 
\begin{eqnarray}
&& \ln Y_{1,2}(v)=
\frac{\pi \beta J}{\cosh\pi v}+
K*\ln \left\{\frac{1+Y_{2,2}}{1+Y_{1,2}^{-1}} \right\}(v)
  \nonumber \\ 
&& \hspace{60pt}
-\ln\left\{\tanh\frac{\pi}{2}(v-x_{1})
            \tanh\frac{\pi}{2}(v+x_{1})\right\}
 \nonumber \\ 
&& \hspace{60pt} 
+\ln\left\{\tanh\frac{\pi}{2}(v-x_{2})
            \tanh\frac{\pi}{2}(v+x_{2})\right\}
  +\pi i,  \nonumber \\ 
&& \ln Y_{m,2}(v)=K*\ln
\left\{\frac{(1+Y_{m+1,2})(1+Y_{m-1,2})}{1+Y_{m,2}^{-1}}\right\}(v)
  \nonumber \\ 
&& \hspace{30pt}  
-\ln\left\{\tanh\frac{\pi}{2}(v-x_{m})
            \tanh\frac{\pi}{2}(v+x_{m})\right\}
 \nonumber \\ 
&& \hspace{30pt} 
+\ln\left\{\tanh\frac{\pi}{2}(v-x_{m+1})
            \tanh\frac{\pi}{2}(v+x_{m+1})\right\}
 \nonumber \\ 
&& \hspace{30pt} 
+\ln\left\{\tanh\frac{\pi}{2}(v-x_{m-1})
            \tanh\frac{\pi}{2}(v+x_{m-1})\right\} 
            +\frac{1+(-1)^{m+1}}{2}\pi i 
 \nonumber \\ 
&&  \hspace{160pt}
  {\rm for} \quad m \in {\mathbb Z}_{\ge 2},
  \label{es-TBA2}   
\end{eqnarray}
where the integral kernel
 is defined in (\ref{kernel}). 
 In addition $x_{m}$ is determined by the 
 condition $Y_{m}(x_{m}\pm i/2)=-1$. 
Through  the above excited state TBA equation (\ref{es-TBA2}), 
the second largest eigenvalue is given as follows: 
\begin{eqnarray}
&& \lim_{N \to \infty}
\ln \left|T_{1,2}^{(1)}
(u_N,0)\right|
=-\beta J\left(\frac{4\pi}{3\sqrt{3}}-1\right)
+2\ln\left(\tanh\frac{\pi}{2}x_{1}\right) \nonumber \\
&&\hspace{3pt}+\int_{-\infty}^
{\infty}G(v)\ln\{(1+Y_{1,2}(v))\tanh\frac{\pi}{2}(v-x_1)
 \tanh\frac{\pi}{2}(v+x_1)\}dv. 
\label{second}
\end{eqnarray}
This second largest eigenvalue (\ref{second}) together with 
the largest one (\ref{largest}) 
describe the correlation length at any finite temperature 
(see, (\ref{corr})). 
 These excited state TBA equation (\ref{es-TBA2}) 
 and Eq. (\ref{second}) will be difficult to obtain
  by the string hypothesis; they exemplify the efficiency of 
  the QTM method. 
\section{Summary and discussion}
We have applied the QTM method to
the $osp(1|2)$ integrable spin chain.
Making use of the $T$-system
and the $Y$-system, we derived the infinite set of
NLIE which provide the free energy  and
the correlation length at any finite temperatures. 
As concerns  the free energy, these NLIE are
identical to the TBA equation based on the string 
hypothesis. 
One can say that the validity of 
the $osp(1|2)$ $T$-system in Ref. \cite{T5} 
has been confirmed 
from the point of view of the QTM method. 
To the author's knowledge, this paper is the first 
explicit derivation of the excited state TBA equation 
for the $osp(1|2)$ spin chain. 

We comment on the ferromagnetic regime $J>0$
in the second largest eigenvalue sector, which
we have not considered in section 5.
In the antiferromagnetic regime $J<0$, the
correlation length are characterized by the
additional zeros of the fusion QTMs.
These zeros are real and symmetric with respect
to the imaginary axis and this symmetry
 will  never break at any finite temperature.
While in the ferromagnetic regime ($J>0$),
the ``level crossing" will occur successively.
This attributes  to the change of  
distribution patterns of the additional zeros
as follows.
First at high temperature, 
two additional zeros are on the imaginary axis
in the physical strip.
However, at low temperature,
this  pattern will 
no longer provide the second largest eigenvalue.

There is another definition of the transfer matrix 
for the $R$-matrix (\ref{rmat2}), which is based on the 
{\em graded} formalism \cite{Kul,KulSk,EK,MR} of the quantum inverse 
scattering method.  In this case, 
the $R$-matrix (\ref{rmat2}) is defined as $R(v)=P^{g}\check{R}(v)$, 
where $P^{g}$ is the {\em graded} permutation operator.
As a result, the ordinary row-to-row transfer matrix is 
defined as the {\em super}-trace of a monodromy matrix, 
and the phase factor (cf.\cite{Mar95-1,Mar95-2})
of the BAE disappears \cite{Kul,MR}.
In the QTM for the graded case, we are 
not sure how the phase factor corresponding 
to the right hand side in (\ref{BAE}) 
changes. 

One might suspect that our TBA equation (\ref{tba}) is 
related to the one \cite{MNTT} for the 
Izergin-Korepin model since the representation space of 
 $osp(1|2)^{(1)}$ has close 
resemblance to that of the affine Lie algebra $A_{2}^{(2)}$. 
However, to our knowledge, 
the TBA equation for the Izergin-Korepin model has been  
constructed \cite{MNTT} so that it reduces to the one for a 
$su(3)$-invariant model in the rational limit. 
%
There will be another branch of the rational limit for 
 the Izergin-Korepin model \cite{Mar95-2}. 
 Together with the above-mentioned graded case, 
 the comparison with our results will be interesting. 
It is an interesting problem to extend a similar analysis 
 to more general $osp(r|2s)$ model. 
Unfortunately, for $r \in {\mathbb Z}_{\ge 2}$ case, 
we have
\footnote{$osp(2|2) \simeq sl(1|2)$ case is an exception to this statement
 \cite{T1,T2,T3,T4,JKSfusion}.}
 only the subset of the $T$-system \cite{T5,T4}. 
While for $r=1$ case, we have already a complete set of 
the $T$-system \cite{T5}, which will be relevant for the QTM method. 
 So we hope to report, at the beginning, 
 the TBA equation for the $osp(1|2s)$ model in the near future. 
Finally we refer to another formulation of NLIE based on 
the  finite number of the auxiliary functions 
(see for example, \cite{Klu,KZeit,JK,JKStJ,FK99,SSSU}). 
 Although there seems to be no connection between 
the TBA equation and such NLIE, 
this formulation will be advantageous to the numerical calculation of 
the physical quantities at finite temperature. 
%
%
\section*{Acknowledgments}
\noindent
ZT would like to thank Professor J. Suzuki for discussions
at the earlier stage of the work \cite{JKSfusion},
which gives a motivation of the present work.
The authors also thank him for his useful comments.
 
They would like to thank Professor A. Kuniba for 
his encouragement. 
This work is supported in part by a Grant-in-Aid for 
JSPS Research Fellows from the Ministry of Education, 
Science, Culture and Sports of Japan. 

\newpage
\listoffigures
\newpage            
\vspace*{1cm}
\noindent
{\bf K.~Sakai and Z.~Tsuboi,}\\
{\bf Thermodynamic Bethe ansatz equation from fusion hierarchy of $osp(1|2)$
integrable spin chain}
\newpage 
\vspace*{1cm}
\noindent
{\bf K.~Sakai and Z.~Tsuboi,}\\
{\bf Thermodynamic Bethe ansatz equation from fusion hierarchy of $osp(1|2)$
integrable spin chain}
\newpage 
\vspace*{1cm}
\noindent
{\bf K.~Sakai and Z.~Tsuboi,}\\
{\bf Thermodynamic Bethe ansatz equation from fusion hierarchy of $osp(1|2)$
integrable spin chain}
\newpage 
\vspace*{1cm}
\noindent
{\bf K.~Sakai and Z.~Tsuboi,}\\
{\bf Thermodynamic Bethe ansatz equation from fusion hierarchy of $osp(1|2)$
integrable spin chain}
\newpage 
\vspace*{1cm}
\noindent
{\bf K.~Sakai and Z.~Tsuboi,}\\
{\bf Thermodynamic Bethe ansatz equation from fusion hierarchy of $osp(1|2)$
integrable spin chain}
\newpage 
%
\vspace*{1cm}
\noindent
{\bf K.~Sakai and Z.~Tsuboi,}\\
{\bf Thermodynamic Bethe ansatz equation from fusion hierarchy of $osp(1|2)$
integrable spin chain} 
\end{document}